\documentclass[a4paper,11pt]{article}
\pdfoutput=1 

\usepackage{jinstpub} 

\title{Streamer studies in Resistive Plate Chambers.}

\author[1]{A. Paoloni,\note{Corresponding author.}}
\author{A. Mengucci,}
\author{M. Spinetti,}
\author{M. Ventura,}
\author{and L. Votano.}
\affiliation{INFN-LNF, \\via E. Fermi 40 00044 Frascati (RM), Italy.}

\emailAdd{alessandro.paoloni@lnf.infn.it}

\abstract{
The present paper is meant as an update of the presentation given in a
previous Resistive Plate Chamber (RPC) workshop, aimed at finding an 
eco-friendly gas mixture for streamer operation of RPCs. 
Indeed the streamer working regime is still 
suitable for building large RPC systems dedicated to low rate applications, 
such as cosmic ray and neutrino physics.
In addition to other studies about gas mixtures for streamer mode operation,
in this paper the replacement of R134a with $CF_4$, a gas widely used in other 
gaseous detectors, has been investigated.

The effect of the gas gap thickness on the discharge quenching has also been 
studied; this is an important check because thin gas gaps of 1 mm, one half of 
the typical used value, have been introduced for high rate applications.

Finally preliminar results about the streamer formation timing are also 
reported. 
}


\proceeding{XIV Workshop on Resistive Plate Chambers and Related Detectors (RPC2018)\\
  19-23 February 2018\\
  Puerto Vallarta, Jalisco State, MEXICO.}

\begin{document}
\maketitle
\flushbottom

\section{Introduction}
\label{sec:intro}
In recent years, in view of upgrades to some of the LHC experiments (ATLAS, 
CMS and ALICE), most of the R\&D activity on Resistive
Plate Chambers (RPCs) has been devoted to avalanche operation (see for instance
\cite{bATLASup}).
Nevertheless, low rate applications, such as neutrino \cite{bopera} or 
astroparticle physics \cite{bargo}, could still use RPCs operated in streamer
mode, profiting of electronics simplicity (no amplification needed) and of the
existence of safe and eco-friendly gas mixtures made of argon (Ar) and 
tetrafluoropropene (HFO-1234ze) \cite{bGent2016}. 

In \cite{bGent2016} results of studies about mixtures containing 
carbon dioxide (CO$_2$), nitrogen (N$_2$), helium (He) and
tetrafluoromethane (CF$_4$), a gas widely used for other gaseous detectors 
(see \cite{bgem} for instance), were also reported. 
In section \ref{sec:cf4} results about additional investigations performed on 
the use of tetrafluoromethane (CF$_4$) are reported.
The effects of the gas gap thickness on the streamer discharge are also
discussed in section \ref{sec:gasgap}; in particular the performance of
RPCs with 1 mm gas gap and 1 mm electrode thickness, a possible new standard
in the detector production for LHC experiments, is shown.
Finally studies about streamer timing are reported in section 
\ref{sec:avastreamer}: these measurements could 
be exploited to disentangle between different models of streamer formation.   

\section{Streamer operation of RPCs filled with gas mixtures containing CF$_4$}
\label{sec:cf4}
Typical gas mixtures for streamer mode operation of RPCs are composed by
Ar and one or more gases, known as ``quenchers'', needed to absorb UV
photons, produced by Ar de-excitation, without re-emission.
Good quenchers are isobutane (i-C$_4$H$_{10}$), whose percentage is limited to 
few \% because of safety requirements on flammability, tetrafluoroethane (R134a)
and tetrafluoropropene (HFO-1234ze, HFO-1234yf).
The addition of sulfur hexafluoride (SF$_6$) permits the operation in
streamer with a strongly decreased charge.
While in \cite{bGent2016} CF$_4$ was investigated as a substitute for
SF$_6$, here it is investigated as a replacement for R134a.

The studies described in this section have been performed, using the same 
set-up, used for gas mixture tests, described in \cite{bGent2016}. 
Cosmic rays crossing the detector under study, a $60\times70$ cm$^2$ wide RPC,
have been triggered by means of scintillators.
The gas gap and the two bakelite electrodes are all 2 mm thick.
Streamer discharges in the gas gap induce opposite polarity signals on both
sides of the chamber.
Those signals are picked up by means of 3.5 cm wide copper strips, faced
to the external faces of the electrodes.
The strips are terminated, on the read-out side, on 110 $\Omega$, and on the
opposite side on their characteristic impedance, 25 $\Omega$.
Signals from the read-out strips are digitized at 5 GS/s and acquired.

\begin{figure}
\centering
\includegraphics[width=14cm]{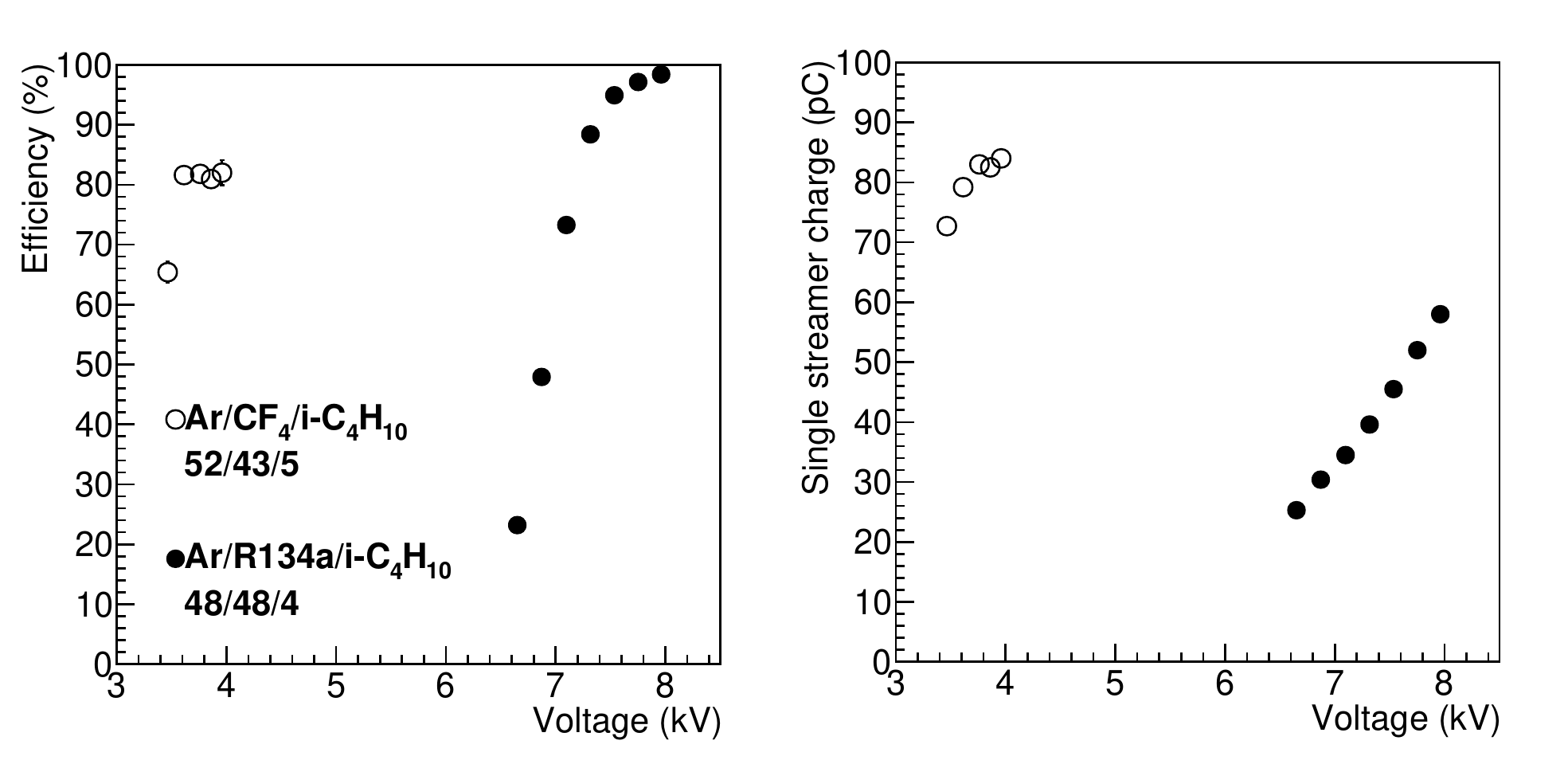}
\caption{Efficiency and single streamer charge, as a function of the operating
voltage, for the gas mixture Ar/CF$_4$/i-C$_4$H$_{10}$ in the volume ratios
52/43/5 (empty circles). 
As a reference, the same measurements performed with the gas mixture 
Ar/R134a/i-C$_4$H$_{10}$ in the volume ratios 48/48/4 (full circles) are also 
reported.}
\label{fig1}
\end{figure}

In figure \ref{fig1} the efficiency and the single streamer charge are shown
as a function of the operating voltage for a gas mixture composed of
Ar, CF$_4$ and i-C$_4$H$_{10}$ in the volume ratios 52/43/5.
As a reference, the same measurements for a gas mixture composed of Ar, 
R134a and i-C$_4$H$_{10}$ in the volume ratios 48/48/4, 
are shown.
The concentrations of Ar and i-C$_4$H$_{10}$ are similar and therefore the
different performances of the two different mixtures can be directly ascribed
to the replacement of R134a by CF$_4$.
The detector, flushed with the mixture containing CF$_4$, reaches the maximum
efficiency at 3.5 kV, more than 4 kV below the reference mixture with R134a; 
the efficiency plateau value is around 80\% with a much higher streamer charge.
This behavior has been observed flushing RPCs with 2 mm gas gap with
mixtures containing small percentages of quencher gases.
As an example, a similar trend was observed in studies about ternary mixtures 
made by Ar, R134a and i-C$_4$H$_{10}$, decreasing the R134a percentage below
20\% at fixed 4\% i-C$_4$H$_{10}$ concentration \cite{bternary}.

\begin{figure}
\centering
\includegraphics[width=7cm]{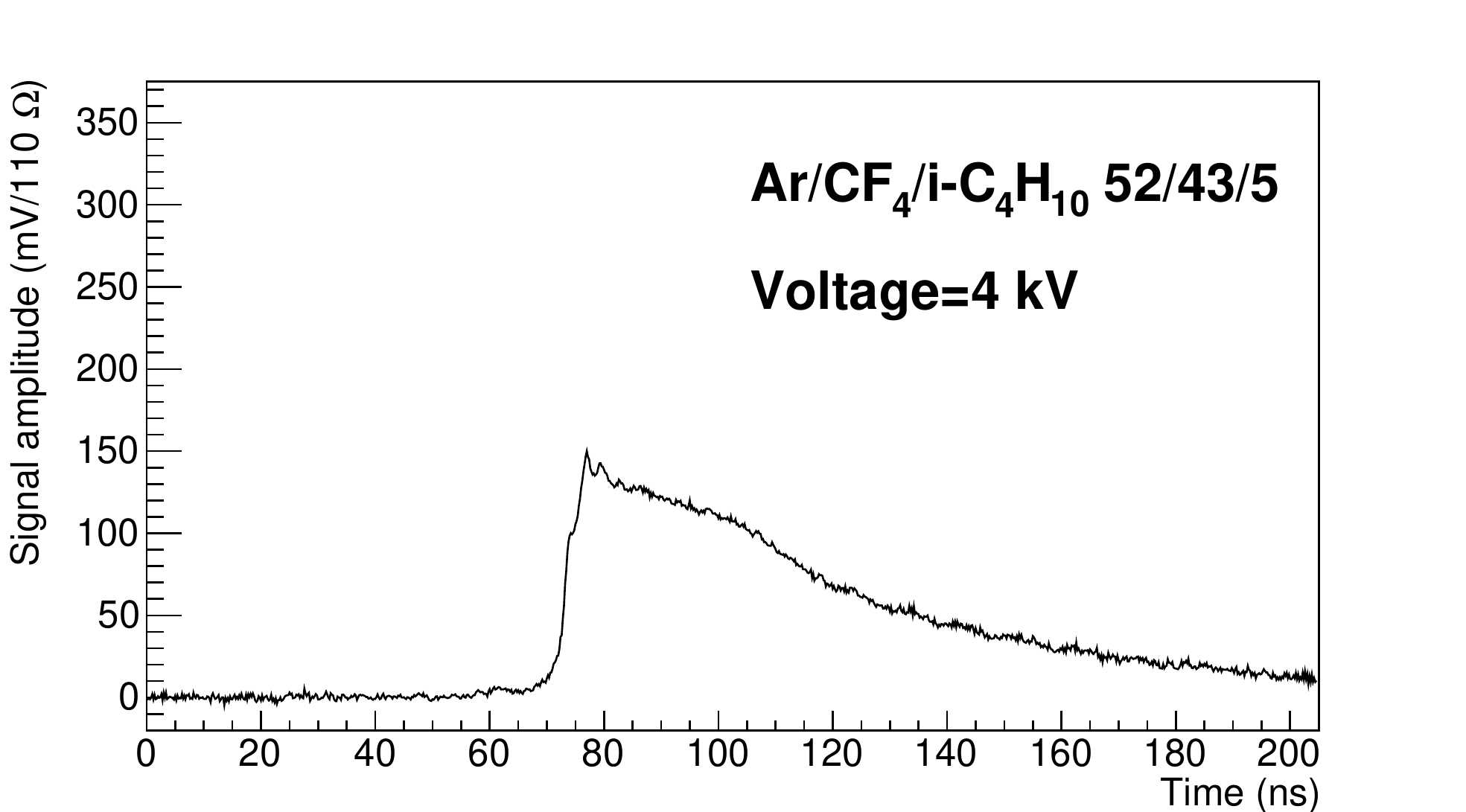}
\includegraphics[width=7cm]{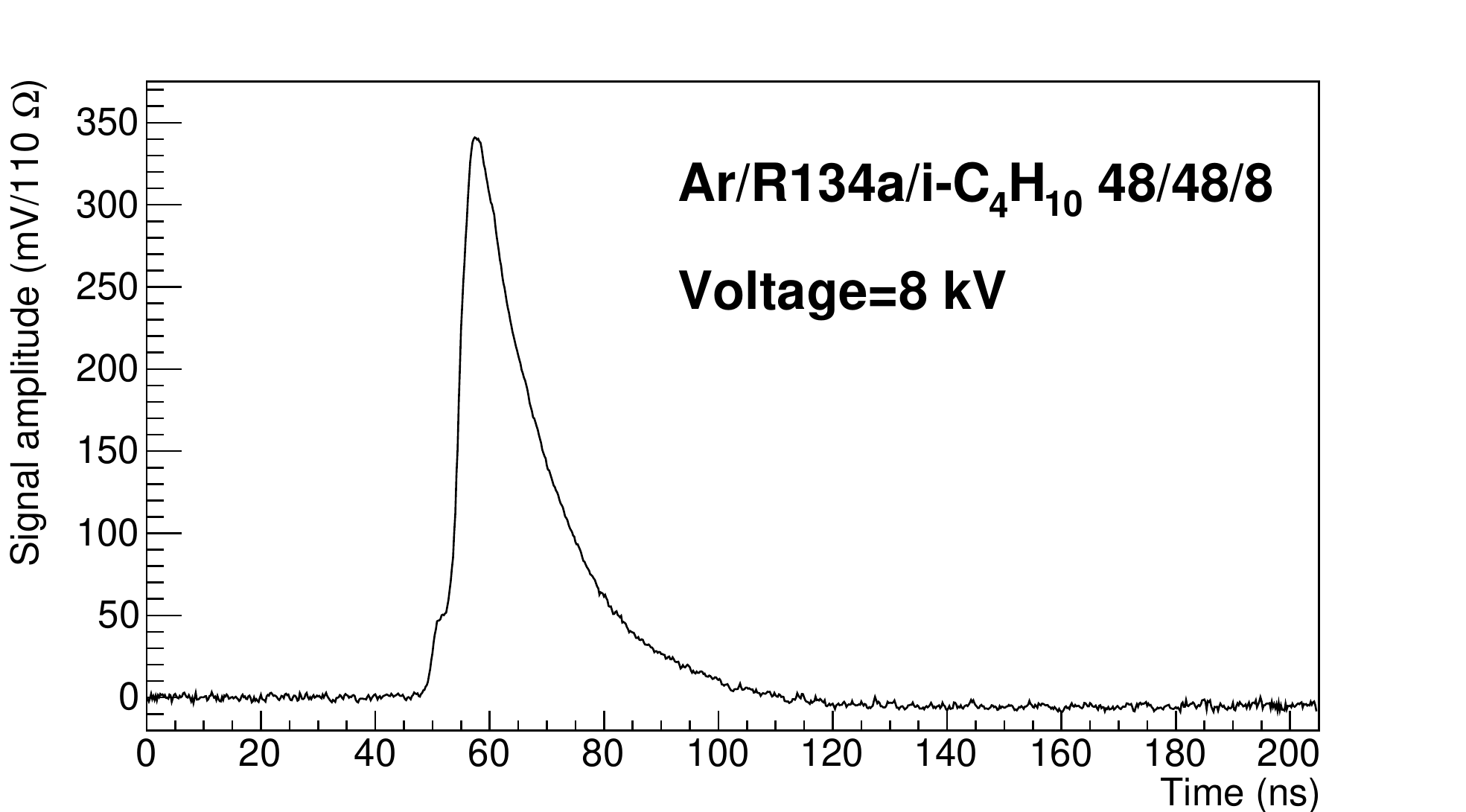}
\caption{Typical wave-forms acquired with the gas mixtures 
Ar/CF$_4$/i-C$_4$H$_{10}$ in the volume ratios 52/43/5 at 4 kV (left) and 
Ar/R134a/i-C$_4$H$_{10}$ in the volume ratios 48/48/4 at 8 kV operating 
voltage (right).}
\label{fig2}
\end{figure}

In figure \ref{fig2} typical single streamer wave-forms are also shown for the
two considered gas mixtures. 
The mixture with CF$_4$ has a higher integrated charge (85 vs 60 pC), resulting
from a lower amplitude (150 vs 350 mV/110 $\Omega$) and a higher 
full-width-at-half-maximum (45 vs 12 ns). 
It has also a higher rise-time (4 vs 2.5 ns from 10\% to 90\% of the signal 
amplitude).

In conclusion CF$_4$ is not advisable as a quencher component inside gas 
mixtures for streamer operation of RPCs, because much better performances,
in terms of efficiency and signal charge, can be obtained with R134a.

\section{Streamer operation of RPCs with 1 mm gas gap and electrode thickness}
\label{sec:gasgap}
Future upgrades of LHC RPC systems, designed for high rate operation in
avalanche mode, are defining a new ``standard'' for the production of bakelite 
RPC gaps.
Instead of 2 mm gas gaps with 2 mm electrode thickness, used so far in large
systems, profiting of the lower charge released in the gas, of the better time
resolution and of the lower operating voltage, future detectors will have
thinner gas gaps.

A study about streamer mode operation of a $10\times10$ cm$^2$ wide
RPC, with 1 mm gas gap and 1 mm bakelite electrodes thickness, is reported in
\cite{bgasgap}.
The detector was read-out by means of a single copper pad and flushed with
gas mixtures made of Ar, HFO-1234ze and SF$_6$.
Decreasing the gas gap, streamers are faster and with lower charge.
The time resolution also improves.

\begin{figure}
\centering
\includegraphics[width=14cm]{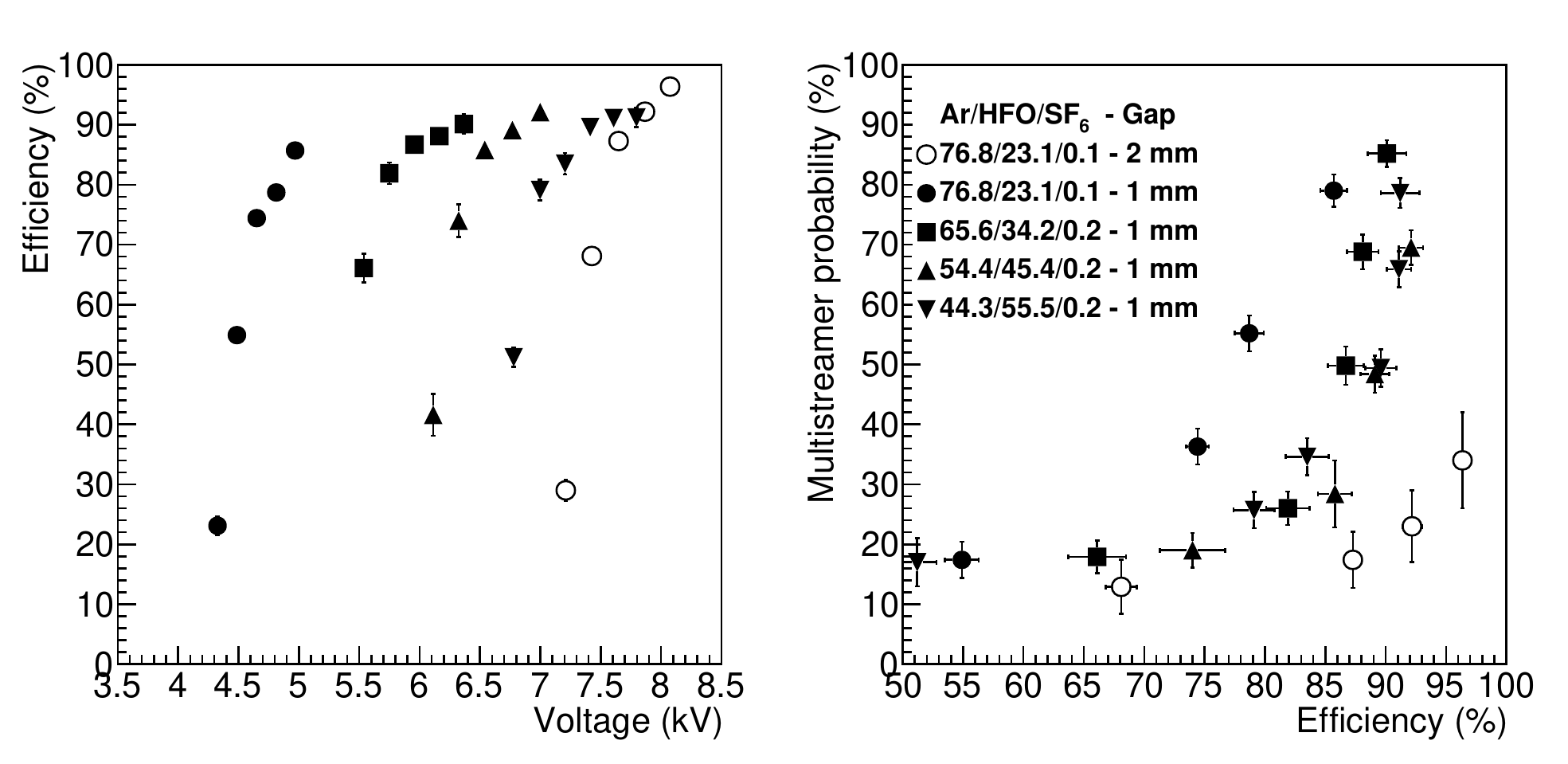}
\caption{Efficiency and multi-streamer probability for streamer operation of
RPCs with gas mixtures made of argon, HFO-1234ze and SF$_6$. Both geometries
with 2 mm (empty markers) and 1 mm (full markers) have been considered.}
\label{fig3}
\end{figure}

In figure \ref{fig3} the measured efficiency values are shown as a function
of the operating voltage. With the considered gas mixtures, smaller plateau
values are obtained using the RPC with 1 mm gas gap.

The multi-streamer probability has been measured as a function of the 
efficiency, checking by eye on a subsample of acquired wave-forms, the
presence of after-pulsing and/or of streamers with an abnormal amplitude
(twice that of the standard ones).
The corresponding results are also shown in figure \ref{fig3}.
It is evident that the multi-streamer probability, at fixed gas gap value,
increases for decreasing HFO-1234ze (the quencher gas) concentrations.
For a fixed gas mixture, the multi-streamer probability strongly increases,
at high efficiency, replacing the 2 mm gas gap RPC with the 1 mm gap one.
In the latter case, above 90\% efficiency, with all the considered gas 
mixtures, the multi-streamer probability is greater than 50\%.
It is therefore preferable not to use 1 mm gas gap RPCs in streamer mode.

\section{Streamer formation time studies}
\label{sec:avastreamer}

\begin{figure}
\centering
\includegraphics[width=15cm]{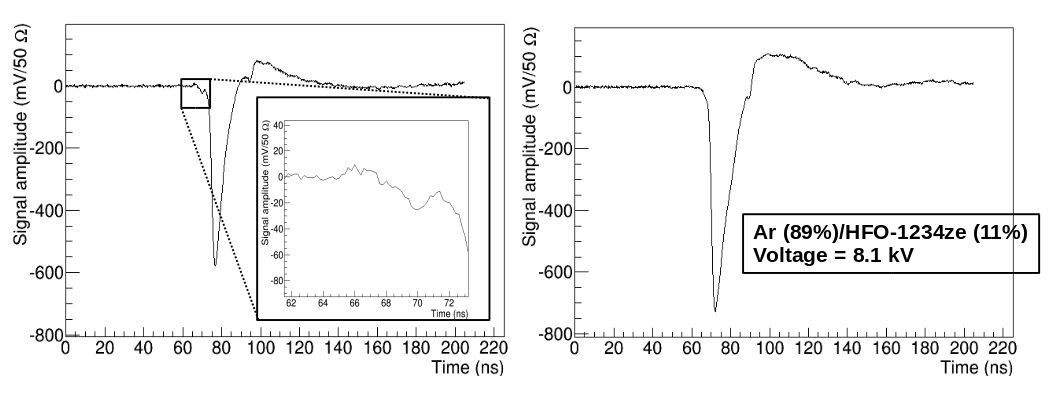}
\caption{Examples of streamer wave-forms acquired at full efficiency for an
RPC detector flushed with a gas mixture composed of Ar (89\%) and HFO-1234ze
(11\%). In the left plot, the avalanche precursor is well separated from the 
streamer and magnified in the insert.}
\label{fig4}
\end{figure}

The time interval between avalanche precursor and streamer
has been measured on a 10$\times$10 cm$^2$ wide RPC with 2 mm gas gap flushed 
with a gas mixture composed of Ar (89\%) and HFO-1234ze (11\%) with and without 
the addition of 0.3\% SF$_6$.
The test has been performed using cosmic rays with a set-up similar 
to that used for the studies of the previous section.
It is worth mentioning that the identification of the avalanche precursor
has not been always possible.
Two acquired wave-forms are shown in figure \ref{fig4} as an example: in the
waveform shown in the left panel, the avalanche precursor is clearly visible 
and separated from the following streamer, while in the waveform of the right
panel it is not.
The streamer delay has been measured only in the first case, as the time 
difference between the minima of the two signals. 

\begin{figure}
\centering
\includegraphics[width=7cm]{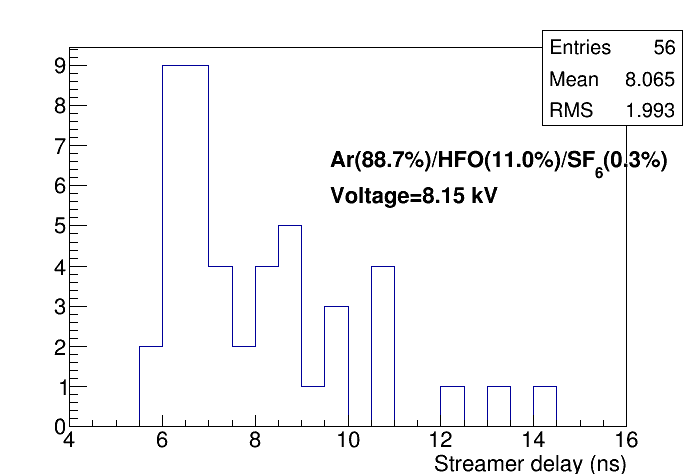}
\includegraphics[width=7cm]{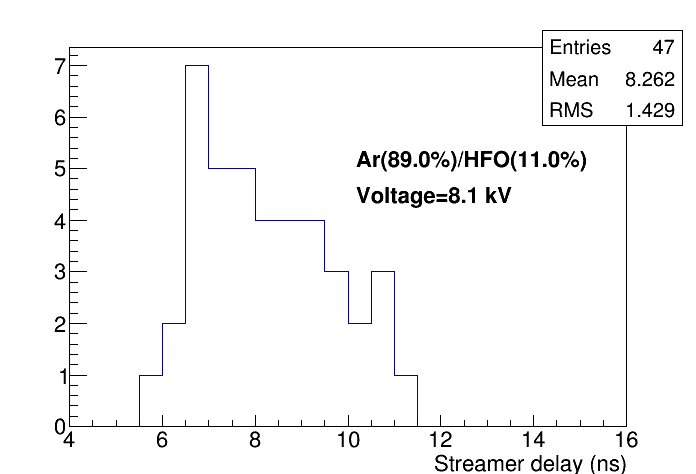}
\caption{Examples of streamer delay distributions measured, with respect to
the avalanche precursor, at full efficiency, for a 2 mm gap RPC flushed with a 
gas mixture composed of Ar (89\%) and HFO-1234ze(11\%) with (left plot) and 
without (right plot) SF$_6$ addition.}
\label{fig5}
\end{figure}

With the considered gas mixtures, the detector reaches full efficiency around
8 kV.
Two examples of the streamer delay distribution, as defined above, are shown 
in figure \ref{fig5} at 8.15 kV (for the mixture containing SF$_6$) and 8.1
kV (for the mixture without SF$_6$) operating voltages, respectively.
The corresponding average values are (8.1 $\pm$ 0.3) ns and (8.3 $\pm$ 0.2)
ns.
The distributions are asymmetrical, with the average value greater than the
most probable one and queues reaching also values few ns above the average 
value.

\begin{figure}
\centering
\includegraphics[width=7cm]{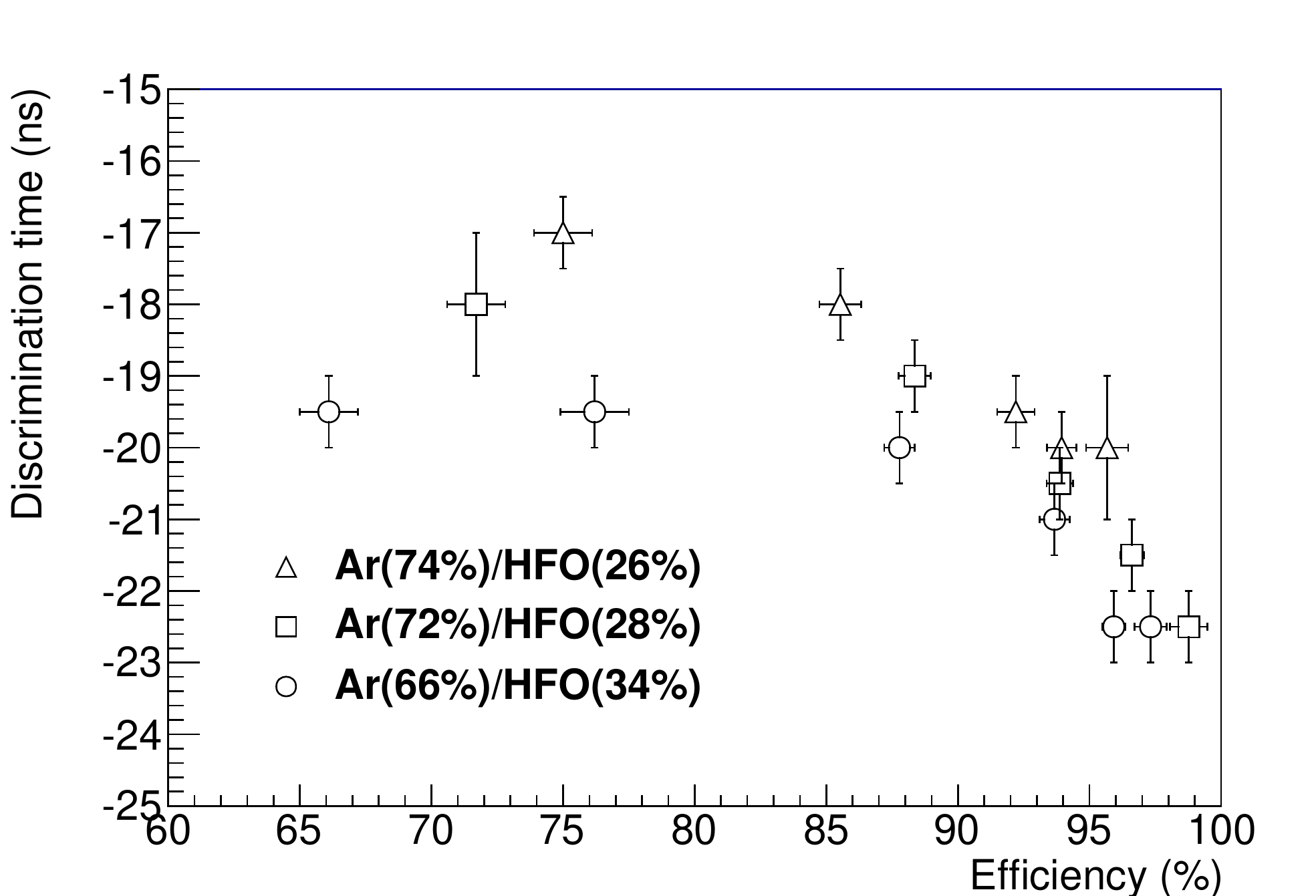}
\includegraphics[width=7cm]{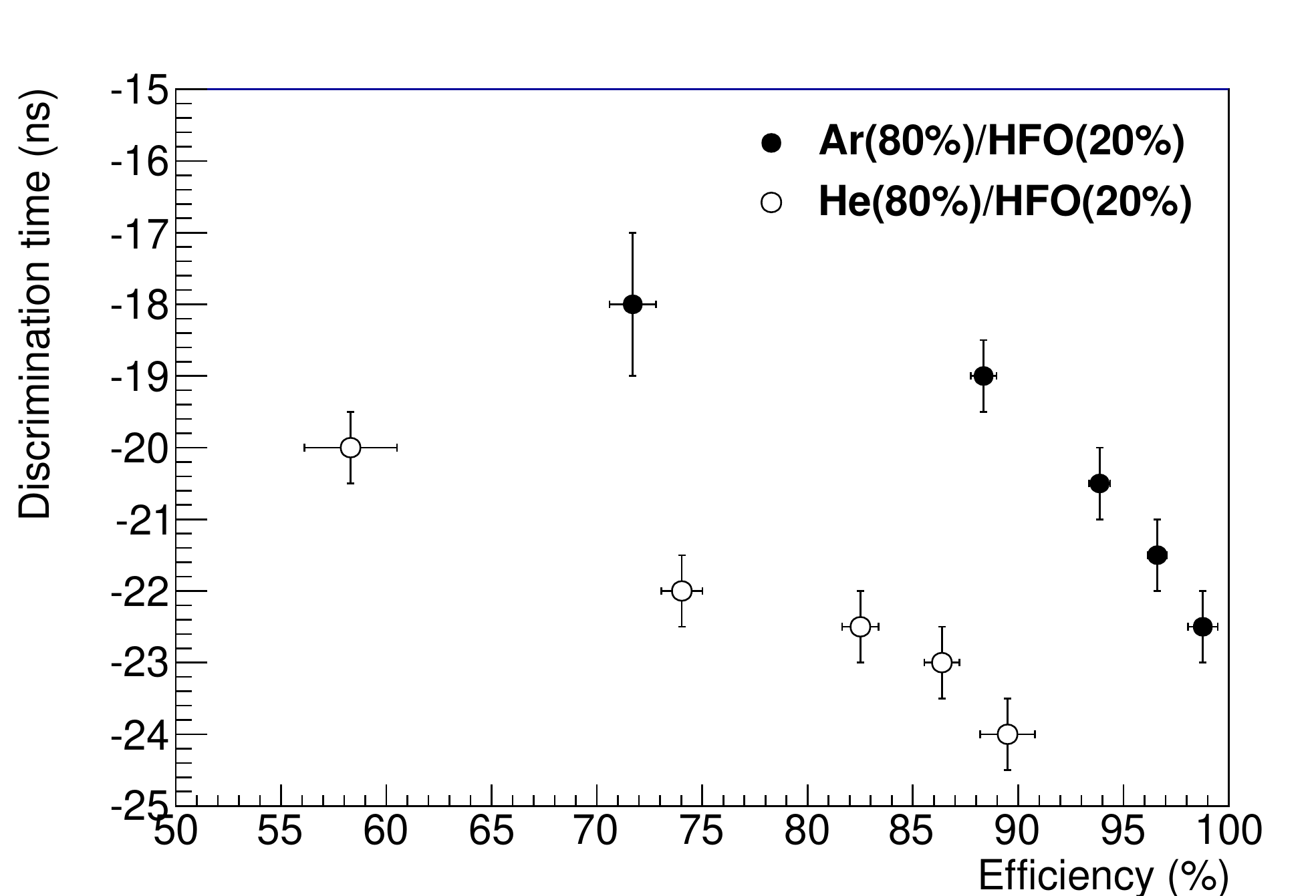}
\caption{Streamer discrimination time vs efficiency for different binary 
mixtures made of Ar and HFO-1234yf (left plot). 
In the right plot a comparison is done at fixed HFO-1234ze concentration 
between Ar and He.}
\label{fig6}
\end{figure}

Other informations about how the streamer timing depends on the gas mixture,
have been obtained profiting of the data collected for the extensive gas 
mixture studies presented in \cite{bGent2016}.
In that analysis, signals, acquired from 3.5 cm wide copper strips, were
discriminated at 50 mV on a 110 $\Omega$ load and the corresponding time
registered.
The discrimination time of the signal from one of the trigger scintillators
was used as a reference and subtracted.
Because of the delays due to the photomultiplier and to the cabling, the
difference between the streamer discrimination and the reference scintillator
times are negative.
In the left panel of figure \ref{fig6}, the streamer discrimination times 
are compared as a function of the efficiency for different HFO-1234yf (the 
quencher component) concentrations in Argon based gas mixtures.
A similar study is performed in the right panel at fixed HFO-1234ze 
concentration, replacing Ar with Helium (He). 

From the plots it is evident that at full efficiency the streamer time
is almost independent from the HFO-1234yf percentage, while the use of He, 
instead of Ar, leads to a faster streamer development.

\section{Conclusions}
\label{sec:conc}
In this paper different aspects of streamer operation of RPCs are discussed.

Concerning gas mixtures studies, tetrafluoromethane has been tested in 
mixtures as a replacement for quencher gases together with argon and isobutane:
results similar to mixtures with insufficient quenching (low operating voltage 
and efficiency, large signals charge and width) have been observed, making
tetrafluoromethane use not advisable in RPCs.

New chambers, built according to the new geometry introduced for avalanche
operation at high rate, with 1 mm gas gap and 1 mm electrode thickness, have
been tested in streamer mode.
A strong increase of the multi-streamer probability is observed, suggesting
to keep the old geometry (2 mm gas gap and electrode thickness) for
streamer mode operation.

Using gas mixtures made of argon and tetrafluoropropene (HFO-1234yf and
HFO-1234ze), streamer timing studies have been performed.
The time separation between avalanche and streamer has been found to be around 
8 ns on average at about 8 kV operating voltage, with distribution tails 
extending to values few ns above.
In addition, a faster streamer formation has been observed, for a fixed
tetrafluoropropene concentration, replacing argon with helium in the gas
mixture.


\end{document}